\documentclass[preprint]{aastex}

\shorttitle{The Variability of T Tau, RY Tau and RW Aur from 1899 to 1952}
\shortauthors{Beck \& Simon}

\begin{document}


\title{The Variability of T Tau, RY Tau and RW Aur from 1899 to 1952}


\author{Tracy L. Beck\altaffilmark{1} and M. Simon\altaffilmark{1}}

\email{tracy@hilo.ess.sunysb.edu}




\altaffiltext{1}{Department of Physics \& Astronomy, SUNY Stony Brook, Stony Brook, NY 11794-3800}



\begin{abstract}

We present the historical light curve of T Tau derived from photographic plates in the Harvard College Observatory archives.  We find that the optical light of T Tau varied by 2-3 (or more) magnitudes on time scales as short as a month prior to $\sim$ 1917, consistent with the results of Lozinskii (1949).  Extreme light fluctuations of greater than 2 magnitudes abruptly ceased in the late 1910's and, to the best of our knowledge, have not repeated since this time.  We compare the observed light variations of T Tau to the T Tauri stars RY Tau and RW Aur, whose light curves we also constructed from inspection of the archival plates.  We find that variable extinction along the line of sight to the star is the most likely explanation for the observed light fluctuation of T Tau during the early part of the 20th century.

\end{abstract}




\section{Introduction}

Joy (1945) identified T Tauri stars as a distinct class of variable stars and Ambartsumian (1947; 1949) first suggested that they represent solar-like stars in the early stages of formation.  They have since that time become understood as such and have been studied vigorously (e.g. Menard \& Bertout 1999 and references therein).

The prototype and one of the brightest members of the class, T Tau, was discovered in the mid 1800's to vary significantly in visual magnitude.   However, T Tau is a complicated system and is not a typical member of the T Tauri class.  Three optical nebulosities are associated with it.  Hind's variable nebula (NGC 1555) and a small nebulosity (NGC 1554) lie $\sim$ 45$''$ and $\sim$ 4$'$ to the west, respectively (Herbig 1950).  The third, Burnham's nebula, extends $\sim$ 5$''$ from the star at a P.A. = 152$^{\circ}$ (Burnham 1890).  Multi-wavelength observations reveal complex distributions of gas and dust associated with the remnant protostellar envelope, molecular outflows (van Langevelde et al. 1994; Momose et al. 1996; Schuster et al. 1997), arcs and filaments of material traced by molecular hydrogen (Herbst et al. 1996; Herbst, Robberto \& Beckwith 1997) and two Herbig-Haro jets (B\"ohm \& Solf 1994; Solf \& B\"ohm 1999).  

T Tau is a young triple system; it consists of T Tau, a K0 star at a V mag $\sim$ 10 (Herbig \& Bell 1988), and an infrared companion (IRC), 0.$''$7 to the south (Dyck, Simon \& Zuckerman 1982), which is itself a binary with a projected separation of 0.$''$05 (Koresko 2000).  We monitored the near infrared flux of the system and found that the optically visible star does not vary significantly in magnitude in the K (2.2 $\mu$m) and L$'$ (3.8 $\mu$m) photometric bands (Beck et al. 2001; Beck 2001).  The IRC binary does vary in near infrared flux on timescales as short as a week.  However, the IRC has never been detected at wavelengths shortward of 1.2 $\mu$m to a limiting V-band magnitude of 19.6 (Stapelfeldt et al. 1998), and therefore does not contribute to the optical variability of the system.

Lozinskii (1949) compiled the historical light curve of T Tau from nearly 2000 observations spanning the interval 1858 to 1941.  A translation of Lozinskii's paper is available online at http://www.ess.sunysb.edu/2001aj/beck01.html.  From 1858 to $\sim$ 1915, T Tau dimmed and brightened randomly in visible light, between 10 and 14 mag in the visible.  After 1915, T Tau appeared mostly bright at $\sim$ 10 mag.  Lozinskii's light curve represents a smoothed 10-day average, and the sources of the individual measurements are unreferenced.  We were therefore interested to use the independent data of the photographic plates available in the Harvard College Observatory archives in order to determine the light curve of T Tau to confirm its large light variations at the begining of the 20th century.  Since the classical T Tauri stars RY Tau and RW Aur appear on the same plates as T Tau, we took advantage of the opportunity to measure their light curves as well.

\section{Data}

We estimated the brightness of T Tau, RY Tau and RW Aur by inspection of over 150 archival AC and AM series photographic plates of the Harvard College Observatory archive.  These patrol plates were obtained using a telescope equipped with a 1.5-inch aperture Cooke lens and provided data with a plate scale of 600$''$/mm spanning the time interval 1899 to 1952.  We supplemented the T Tau data with more than 125 measurements from the plates in the RH and RB series, which were obtained using 3-inch aperture Ross lens at a scale of 390$''$/mm and covered the period 1928 to 1952.

The spectral response of the photographic plates was essentially blue, so we estimated the magnitude of T Tau, RY Tau and RW Aur by comparison with the B magnitudes of nearby stars.  The estimated magnitudes of our targets are therefore B magnitudes in this relative sense.  We first obtained the B-band magnitudes for three dozen stars within 1$^{\circ}$ of T Tau using the SIMBAD database and USNO catalog (Monet et al. 1996).  By identifiying these reference stars on plates spanning several decades we eliminated the obvious variables among them and determined the limiting sensitivities of the AC/AM and RH/RB plates to be 12-13.5 mag and 14-15 mag, respectively.  We estimated the magnitude of T Tau, RY Tau and RW Aur with reference to 16 stars with B magnitudes ranging from 9.6 to 13.8 in increments of 0.2 to 0.3 magnitudes.  The internal precision of the estimated magnitudes of the targets is $\sim$$\pm$0.3 mag and is determined by the magnitude spacing of the calibrators and the accuracy of their individual B-band magnitudes.  Figures 1 through 3 present the derived light curves for T Tau, RY Tau and RW Aur. A table of the data used to generate these light curves is available online at the webpage referenced in $\S$ 1.  Inspection of Digitized Sky Survey images revealed that the target stars are the brightest objects within a 3$'$ radius, hence we attribute the observed light fluctuations solely to them.




\section{Results}

The light curve presented in Figure 1 is similar to that determined by Lozinskii (1949) in that the magnitude of T Tau dimmed and brightened randomly in the early part of the 20th century from B $\sim$ 11 to fainter than the detection limit on timescales as short as a month.  To demonstrate the time scale of the observed light variations, Figure 4 presents 15 measurements of the magnitude of T Tau during 1902.  From 10 January to 10 February 1902 T Tau brightened from 13 mag to $\sim$11 mag, and six measurements in October-November 1902 show that it varied by $\sim$1 mag on time scales of less than two weeks.  A Lomb Normalized Periodogram analysis (Press et al. 1994) of the light curve between 1899 and 1917 shows no significant periodicity on time scales of less than $\sim$ 5 years.   From the late 1910's to 1952, T Tau stabilized at a B magnitude of $\sim$ 11 and, aside from brief dimming events of 1.5 magnitudes in 1925 and 0.8 magnitudes in 1931, did not vary further at a statistically significant level.   Additional data on the variability of T Tau, obtained from the online database of Herbst et al. (1994, hereafter H94) and the American Association of Variable Star Observers (AAVSO; Mattei 2000), shows that between 1937 and the present it has not varied by more than $\sim$ $\pm$0.5 magnitudes from its average.  Observations of the optical variability of T Tau taken during a single week reveal that it presently does not vary by more than $\sim$ 0.1 V-band magnitudes on this time scale (Ismailov 1997).  The ``flickering'' of 2-3 magnitudes observed in T Tau in the early years of 20th century was real and ceased abruptly in $\sim$ 1917.  

The light curve of RY Tau (Figure 2) is characterized by two time scales.  It varies by 2-3 magnitudes over approximately a decade and by $\sim$ 1 magnitude on time scales of less than a year.  This type of variability has continued to the present (H94; Holtzman, Herbst \& Booth 1986; Petrov et al. 1998).  RW Aur varied by 2-3 magnitudes on timescales as short as a month throughout the 1899-1952 interval (Figure 3).  H94 report similar variability; apparently this behavior has continued for at least a century.  Although the amplitudes and time scales of the light fluctuations of RW Aur and T Tau before 1917 are similar, we argue in $\S$4 that their variability is probably caused by different mechanisms.

\section{Discussion}

Parenago (1954) described a classification which is useful to categorize the light curves of T Tauri stars (see also Herbig 1962):

Class I - The star is more often bright than dim; its brightness is most frequently found in the brighter half of the observed magnitude range.

Class II - The fluctuations are generally close to the middle of the observed range in magnitude.

Class III - The star is more often dim than bright; its brightness is most frequently found in the fainter half of the observed magnitude range.

Class IV - The stellar brightness is observed at all levels of its magnitude range; it is not more likely to be found in a bright, faint or average state.

Based on this scheme, T Tau was a Class IV variable from 1899 to 1917, and a Class I from 1917 to the present.  RY Tau and RW Aur are Class IV variables during the interval we have sampled, although breaking the RY Tau data into sections, as in Figure 2, suggests that its variability may be migrating from one class to another on decade-long time scales. 

Parenago's classification is useful because it draws attention to the fact that the character of T Tau's variability changed significantly and abruptly.  It does not, however, provide interpretation of the causes of the variability.  In their study of the light and color variability of $\sim$ 80 T Tauri stars, H94 argued that the variability of most T Tauris is produced by one, or a combination, of three distinct mechanisms:

Type I - Rotational modulation by cool star spots.  This type of variability is periodic.

Type II - Variable accretion rate or the rotation modulation of accretion hot spots.  This type of variability can be periodic or random.

Type III - Random variations of light which are likely produced by variable obscuration along the line of sight to the star (Herbst \& Shevchenko 1999, Bertout 2000)

Photometric and spectroscopic monitoring studies show that the optical variability of RY Tau is likely caused by variable obscuration (Holtzman, Herbst \& Booth 1986; Petrov et al. 1998), thus it is best described as a Type III variable.  H94 argued that T Tau is the only star in their sample that shows characteristics of all three variability mechanisms.  Herbst et al. (1986) discovered rotational modulation of T Tau's light with a period of 2.8 days. The amplitude of the rotational modulation is small, $\sim$ 0.02 mag, so it seems very unlikely that this modulatation could account for the 2-3 magnitude fluctuations in the early 1900's.

The results of Herbst et al. (1982) and Basri \& Bathala (1990) show that the light fluctuations of RW Aur correlate with H$\alpha$ and veiling emission.  The variability of RW Aur is therefore best classified as Type II (H94).  Although the amplitude and timescale of the light variations of T Tau in the early years of the 20th century are similar to those of RW Aur, it is improbable that variable accretion was responsible for T Tau's variability.  It is very difficult to imagine how the accretion rate, after varying significantly enough to produce light fluctuations of 2-3 magnitudes, could stabilize in a high state with a sufficiently constant amplitude to produce light variations of less 0.5 magnitudes around B $\sim$ 11.
 
The most likely explanation for the observed variability of T Tau prior to 1917 is variable extinction along the line of sight to the star.  T Tau is now  obscured by A$_{v}$ $\sim$ 1 (Cohen \& Kuhi 1979), but clumpy material with 2-3 magnitudes of extinction moving through the line of sight to the star could explain the observed dimmings.  The eventual clearing of this material would result in a more or less constant light output.  Hogerheijde et al (1997) and Akeson et al. (1998) detect a circumstellar disk around T Tau by its emission at mm wavelengths and conclude that it is observed nearly face-on.  Variations in extinction toward T Tau are therefore not likely to be caused by material in its own disk, unless it is strongly warped.

Weintraub et al. (1989) found evidence for Keplerian rotation in material surrounding T Tau, and Momose et al. (1996), Schuster et al. (1997), and Weintraub et al. (1999) detected complex distributions of material associated with the T Tau system.  Two Herbig-Haro flows are associated with T Tau; HH 155 originates from the optically visible star, and HH 355 from the IRC  (B\"ohm \& Solf 1994; Herbst, Robberto \& Beckwith 1997; Solf \& B\"ohm 1999).   Hence, it is conceivable that material associated with the envelope or outflows in the T Tau system could have caused the observed extinction variability of T Tau.  The ``flickering'' and abrupt clearing of the obscuring material implies significant density inhomogeneities on spatial scales of a few stellar radii.  It is reasonable to speculate that the possible variation in extinction presently observed toward the T Tau IRC (Beck, Prato \& Simon 2001) is caused by a similar distribution of material responsible for the optical variability of T Tau in the early 1900's.

In a region as complex as that of T Tau, it seems impossible to identify now the material that would have produced $\sim$3 magnitudes of extinction a century ago.  A reasonable scenario for the variable obscuration must also explain the abrupt cessation and apparent non-repeatability of the variability.  It is worth noting that the components of the T Tau triple system were likely positioned differently a century ago.  Roddier et al. (2000) and Ghez et al. (1995) monitored the orbital motion of the T Tau North-South system from 1989 to 1999 and found that the position angle of T Tau S has changed by $\sim$6$^{\circ}$ with respect to T Tau N, but that it has not varied significantly in angular separation ($<$ 0.$''$03).  Hence, the period of the T Tau N-S system, at a 100 AU separation, is likely on the order of $\sim$500-700 years.   The variability may have been the result of the different geometry of obscuring material associated with the disks or outflows of the system components.  Like T Tau's historic variability, we speculate that other puzzling aspects of T Tauri behavior, such as the discovery of the IRC to UY Aur when it was bright in the visible (Joy \& van Biesbroeck 1944), may have extrinsic rather than intrinsic causes.  A more complete understanding of the orbital motion and orientation of the disks and outflows of these systems will enable tests of this idea.

\acknowledgments

We thank G. Herbig for conversations that led to this study and for valuable advice in its course.  We also thank M. Hazen and A. Doane for their assistance and hostpitality during our use of the Harvard College Observatory plate archives, W. Herbst for providing the database of T Tauri star variability, W. Sherry for his assistance locating potential calibration stars, S. Wolk for his advice regarding periodogram analysis techniques, D. Beznosko for help with the translation of Lozinskii's article, and the referee for a helpful report.  We have used, and acknowledge with thanks, data from the AAVSO International Database, based on observations submitted to the AAVSO by variable star observers worldwide.  This work also made use of the SIMBAD database, operated at CDS. Strasbourg, France, and the Digitized Sky Surveys which were produced at the Space Telescope Science Institute under U.S. Government Grant NAG W-2166.  Our work was supported in part by NSF Grant 98-19694.



\clearpage

\clearpage

\appendix






\clearpage



\figcaption[figure1.ps]{The light curve of T Tau derived by inspection of over 275 archival plates of the Harvard College Observatory.}

\figcaption[figure2.ps]{The light curve of RY Tau derived by inspection of over 150 archival plates of the Harvard College Observatory.}

\figcaption[figure3.ps]{The light curve of RW Aur derived by inspection of over 150 archival plates of the Harvard College Observatory.}

\figcaption[ttaufigure4.ps]{The observed variability of T Tau from 15 archival plate fields observed in 1902.  Significant variations of $\sim$2 magnitudes occurred on time scales of a single month.  The typical uncertainties are $\pm$0.3 mag.}

\begin{figure}
\plotone{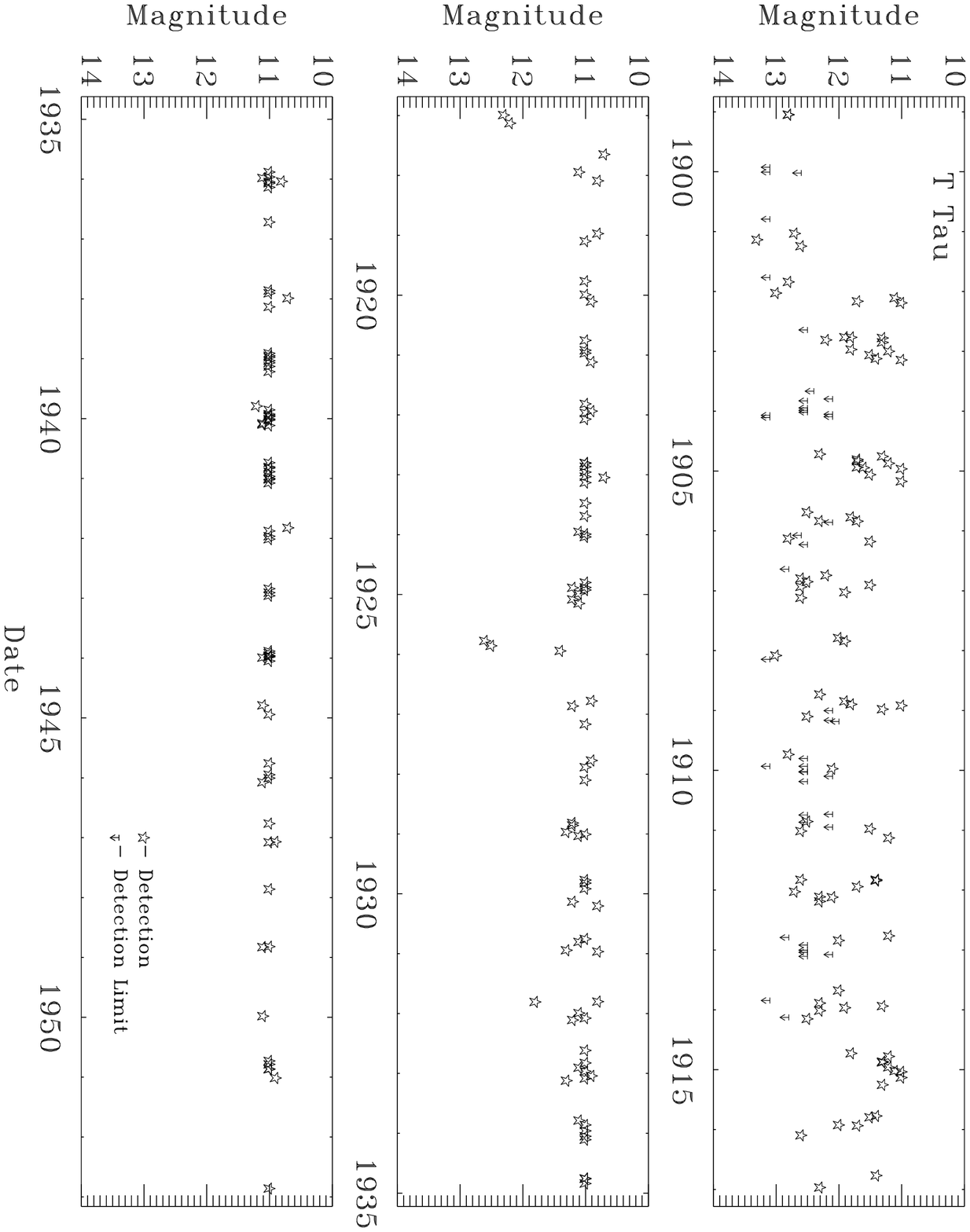}
\end{figure}

\clearpage

\begin{figure}
\plotone{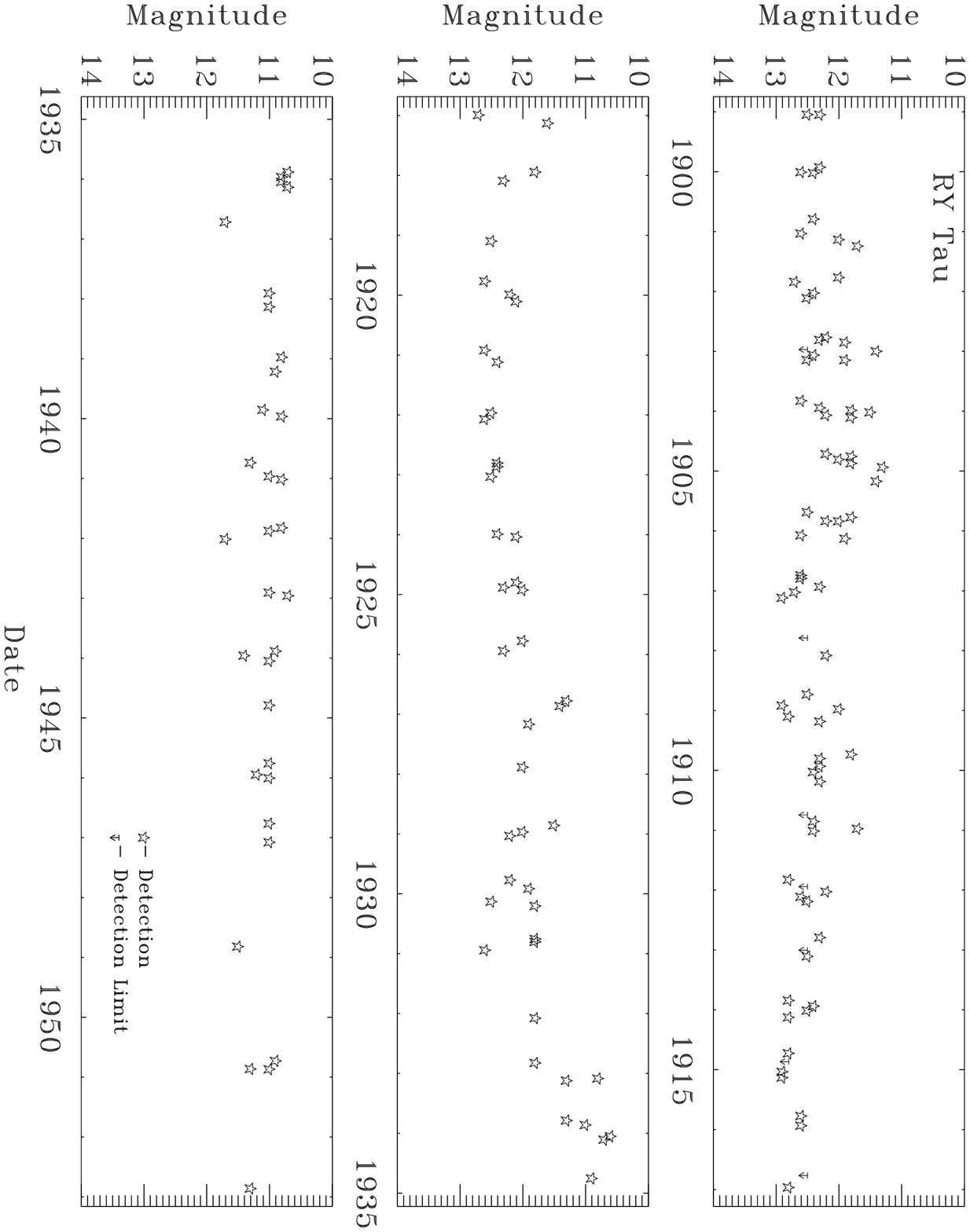}
\end{figure}

\clearpage

\begin{figure}
\plotone{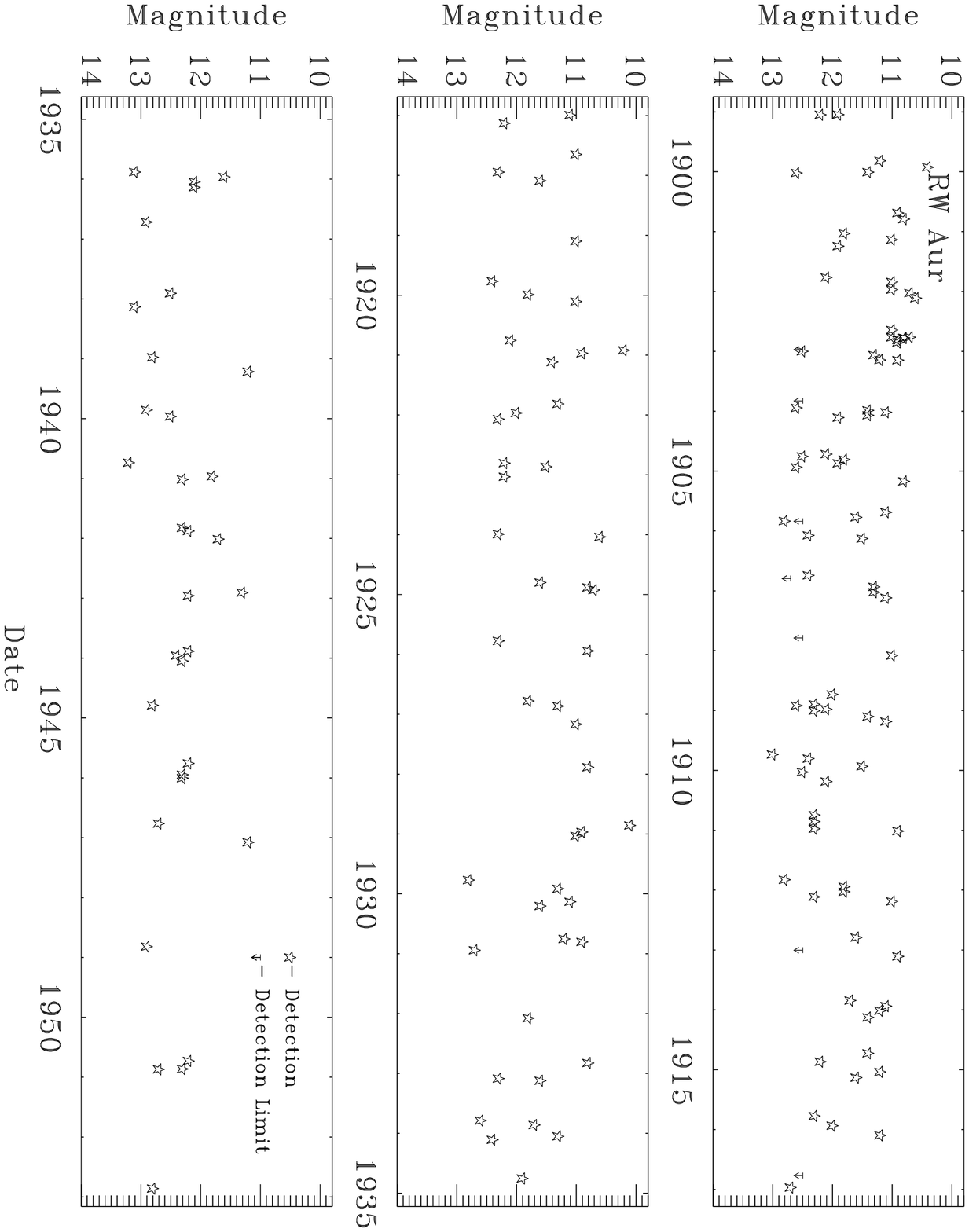}
\end{figure}

\clearpage

\begin{figure}
\plotone{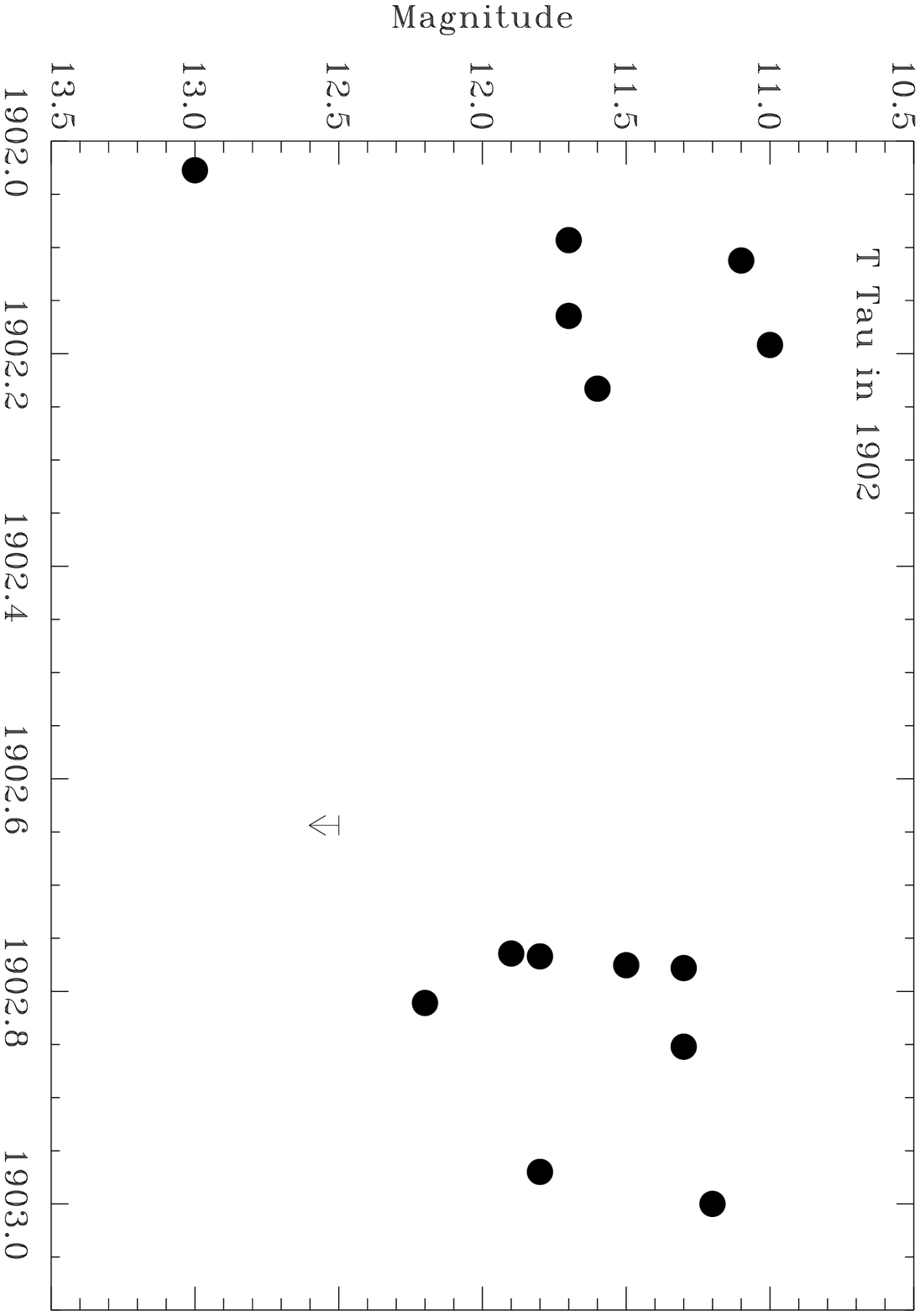}
\end{figure}



\end{document}